\documentclass[journal]{IEEEtran}
\ifCLASSINFOpdf
\else
\fi
\usepackage{amsmath}
\usepackage{amssymb}
\usepackage[pdftex]{graphicx}
\usepackage{algorithm}
\usepackage{algorithmicx}

\usepackage{algpseudocode}
\usepackage{color}
\usepackage{multirow}
\usepackage{makecell}

\usepackage{cite}
\usepackage[hidelinks]{hyperref}

\usepackage{subfigure}
\usepackage{threeparttable}
\usepackage{stfloats}

\newtheorem{remark}{Remark}

\begin{document}
%
\title{Residual Deep Reinforcement Learning  for Inverter-based Volt-Var Control}
%
%

\author{Qiong~Liu,
        Ye~Guo,
        Lirong~Deng,
        Haotian~Liu,
        Dongyu~Li,
        and Hongbin~Sun
\thanks{This work was supported by the National Key R\&D Program of China (2020YFB0906000, 2020YFB0906005).}
\thanks{Qiong Liu, Ye Guo are with the Tsinghua-Berkeley Shenzhen Institute, Tsinghua University, Shenzhen, 518071, Guangdong, China, e-mail: guo-ye@sz.tsinghua.edu.cn.}
\thanks{Lirong Deng is with the Department of Electrical Engineering, Shanghai University of Electric Power, Shanghai, 200000, China}
\thanks{Haotian Liu, Hongbin Sun are with the State Key Laboratory of Power Systems, Department of Electrical Engineering, Tsinghua University, Beijing 100084, China}
\thanks{Donyu Li is with the School of Cyber Science and Technology, Beihang University, Beijing, 100191, China.}
}

\maketitle

\begin{abstract}

A residual deep reinforcement learning (RDRL) approach is proposed by integrating DRL with model-based optimization for inverter-based volt-var control in active distribution networks when the accurate power flow model is unknown.
RDRL learns a residual action with a reduced residual action space, based on the action of the model-based approach with an approximate model.
RDRL inherits the control capability of the approximate-model-based optimization and enhances the policy optimization capability by residual policy learning. Additionally, it improves the approximation accuracy of the critic and reduces the search difficulties of the actor by reducing residual action space.
To address the issues of ``too small" or ``too large" residual action space of RDRL and further improve the optimization performance, we extend RDRL to a boosting RDRL approach. It selects a much smaller residual action space and learns a residual policy by using the policy of RDRL as a base policy.
Simulations demonstrate that RDRL and boosting RDRL improve the optimization performance considerably throughout the learning stage and verify their rationales point-by-point, including 1) inheriting the capability of the approximate model-based optimization, 2) residual policy learning, and 3) learning in a reduced action space. 
\end{abstract}


%
\IEEEpeerreviewmaketitle

\section{Introduction}

\IEEEPARstart {T}o achieve a carbon-neutral society, more distributed generations (DGs) will be integrated into active distribution networks (ADNs). 
The high penetration of DGs may cause severe voltage problems. 
Recently, most DGs are inverter-based, which can provide reactive power rapidly with step-less regulation. Inverter-based Volt-Var control (IB-VVC) has attracted increasing interest.

Model-based optimization methods are widely used to solve IB-VVC problems  \cite{farivarInverterVARControl2011, zhangHierarchicallyCoordinatedVoltageVAR2020}.
Those methods can obtain a reliable solution under the accurate power flow model of ADNs.
However, in real applications, it may be difficult to obtain a high-accuracy model for the distribution system operator due to the complex structure of distribution networks \cite{albertStructuralVulnerabilityNorth2004}. 
The control performance deteriorates with the decreased model accuracy.

As a model-free method, deep reinforcement learning (DRL) has made breakthrough achievements in computer games, chess go, robots, and self-driven cars, which also attract huge interest in VVC problems \cite{chenReinforcementLearningSelective2022,liuBiLevelOffPolicyReinforcement2023}. 
For VVC problems, DRL has two attractive advantages:
1) It learns to make actions from interactive data, and a precise model is not needed;
2) It has a high computation efficiency that only needs a forward computation of the neural network in the application stage because the time-consuming optimization process is shifted into the training stage.
However, DRL also suffers from optimality and convergence issues. 
During the early stages of learning, a DRL agent may experience poor VVC performance due to the lack of training \cite{liuTwoStageDeepReinforcement2021}. Even after sufficient training, a small optimal gap may still exist due to the estimation error of neural networks.
Existing efforts on improving optimality and convergence can be categorized into three types.


First is improving the reward function by trading off the weight of power loss and voltage violation.
A small penalty factor of voltage violation cannot penalize the voltages into the normal range, whereas a large factor results in worse or even unstable learning performance \cite{zhangDeepReinforcementLearning2021, wangSafeOffPolicyDeep2020}.
To alleviate the problem, paper \cite{zhangDeepReinforcementLearning2021} uses a switch reward method to give priority to eliminating voltage violations. If voltage violations appear, the reward only contains the penalty of voltage violation.
Paper \cite{wangSafeOffPolicyDeep2020} designs a constrained soft actor-critic algorithm to tune the ratio automatically.
A well-designed reward function can speed up the convergence of DRL and enhance the VVC performance.
Nevertheless, it does not solve the issue of weak VVC performance during the initial learning stage, and there is still room for improvement in VVC performance after enough time to learn.


Second is selecting a suitable DRL algorithm or making specific modifications according to the characteristics of the VVC problem. 
It is difficult to find the "best" DRL algorithm for all tasks. For example, the original paper on soft actor-critic (SAC) only shows SAC outperforms other algorithms in 4 out of 6 tasks \cite{haarnojaSoftActorCriticOffPolicy2018}. Paper \cite{wangSafeOffPolicyDeep2020} also shows that a constrained soft actor-critic has a faster convergence process and better optimality compared with a constrained proximal policy optimization algorithm. Graph neural networks can be introduced to DRL to improve the robustness and filter noise \cite{yanMultiAgentSafe2023, caoPhysicsinformedGraphicalRepresentationenabled2023}. 
However, relying solely on DRL approaches has not successfully addressed the issues of optimality and convergence. 

Third is utilizing a power flow model to assist DRL.
A two-stage DRL framework first trains a robust policy in an inaccurate power flow model and then continually improves the VVC performance by transferring in a real environment \cite{liuTwoStageDeepReinforcement2021}. It alleviates weak performance during the initial learning stage when training DRL on a real ADN directly.
To improve the data efficiency, model-based DRL learns a power flow model from historical operation data and then trains the policy on the learned model \cite{gaoModelaugmentedSafeReinforcement2022, liuRobustOfflineDeep2021}.
To ensure no voltage violation appears in the training process, an optimization algorithm based on the approximated power flow model is designed to readjust the action of DRL when DRL makes unsafe actions \cite{kouSafeDeepReinforcement2020,gaoModelaugmentedSafeReinforcement2022}.
Those approaches mainly focus on the safety issues of DRL, and further research is needed to improve optimization capabilities by utilizing a power flow model to assist DRL.

As discussed above, the DRL-based VVC performance can be improved from three perspectives: design better reward functions, select or modify suitable DRL algorithms, and utilize the power flow model to assist DRL. 
However, the two problems still have not been addressed successfully:
1) the optimization capability in the initial learning stage is weak, and 2) there is still a small optimal gap after enough time to learn.
To alleviate the above two problems, we 
utilize a residual DRL (RDRL) approach \cite{silverResidualPolicyLearning2018, johanninkResidualReinforcementLearning2019,zhangResidualPolicyLearning2022} that integrates model-based optimization with DRL. As shown in Fig. \ref{RDRL_overview}, RDRL learns a residual action with a reduced residual action space on top of a model-based optimization with an approximate model. 
We also extend RDRL to a boosting RDRL (BRDRL) approach that improves the RDRL performance by learning a residual policy on top of the policy of RDRL in a further reduced residual action space.
Compared with the existing literature \cite{liuTwoStageDeepReinforcement2021,  zhangDeepReinforcementLearning2021,  wangSafeOffPolicyDeep2020, 
gaoModelaugmentedSafeReinforcement2022,  liuRobustOfflineDeep2021,  kouSafeDeepReinforcement2020}, the main contributions of this paper are the following:

\begin{figure}[!t] 
\centering
\includegraphics[width=3in]{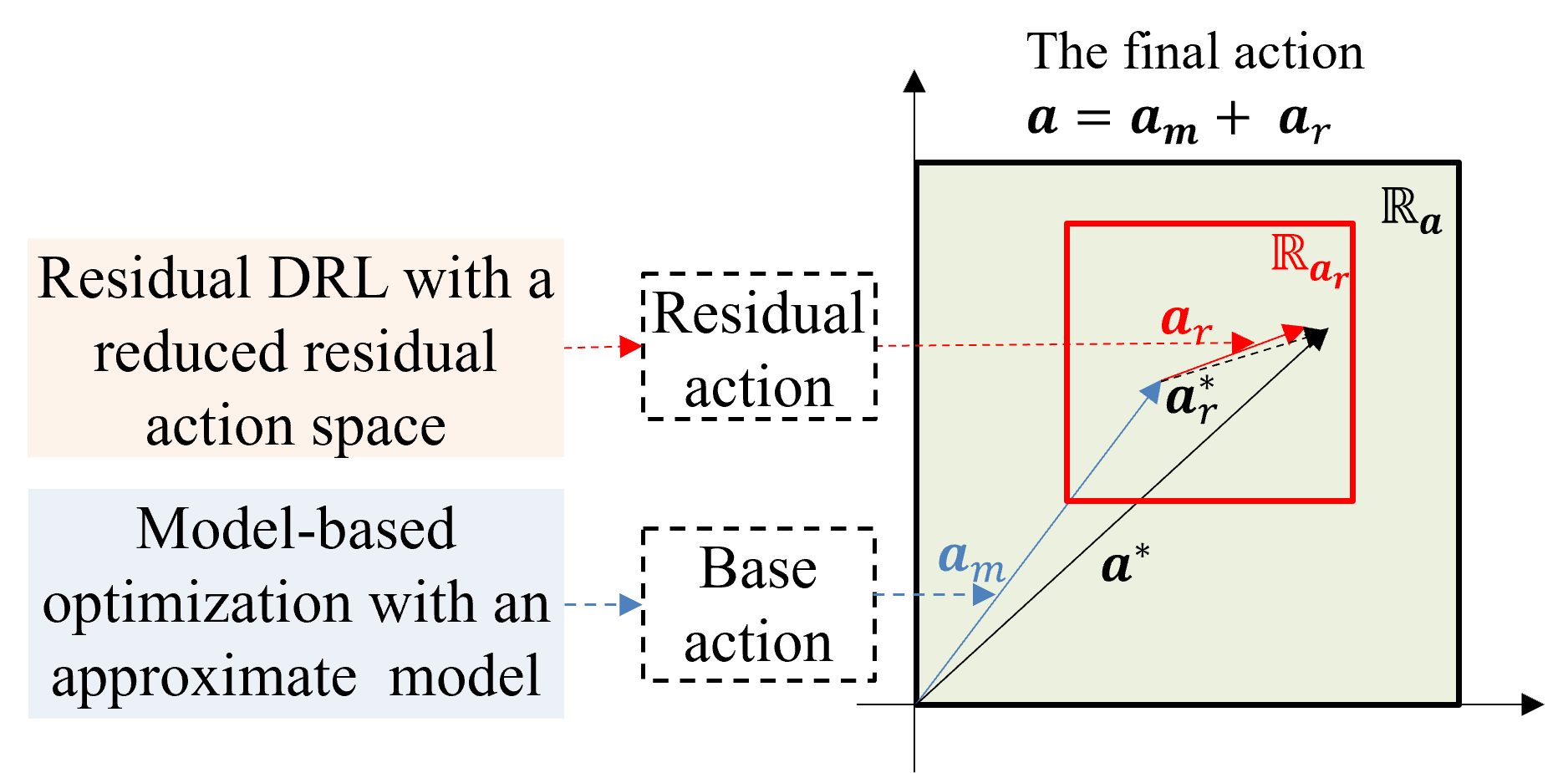}
\caption{Overall structure of the proposed residual DRL framework. $\mathbb{R}_a$ is the original action space, 
$\mathbb{R}_{a_r}$ is the residual action space, 
$a^*$ is the optimal action,
$a_m$ is the action of model-based optimization with an approximate model,
$a_r^*$ is the optimal residual action, and 
$a_r$ is the residual action of residual DRL 
 }
\label{RDRL_overview}
\end{figure}

\begin{itemize}
\item[1.] RDRL learns a residual action of the model-based optimization with an approximate model. It inherits the capabilities of the model-based optimization approach and improves the policy optimization capability by residual policy learning. 



\item[2.] RDRL learns in a reduced residual action space, which alleviates the search difficulties of the actor. Moreover, all generated actions in a reduced action space alleviate the approximation difficulties of the critic, thus improving the approximation accuracy of the critic.

\item[3.] BRDRL improves the optimization performance further based on RDRL. It alleviates the problems of ``too small" or ``too large" action space of RDRL.
 \end{itemize}

The remainder of the paper is organized as follows. Section II introduces the problem formulation of IB-VVC. Section III proposes 
an RDRL approach and designs a residual soft actor-critic algorithm. To improve the performance of RDRL, a BRDRL is designed in 
section IV.
Section V verifies the superiorities of the proposed RDRL and BRDRL through extensive simulations. Section VI concludes the results.

\section{Problem Formulation}
This section introduces the formulations of model-based IB-VVC and DRL-based IB-VVC.

\subsection{Inverter-based Volt-Var Control}
IB-VVC minimizes power loss and eliminates the voltage violation of ADNs by optimizing the outputs of inverter-based devices. It is usually formulated as a constrained optimal power flow \cite{farivarInverterVARControl2011, zhangHierarchicallyCoordinatedVoltageVAR2020}. For generality, we use the simplified version adopted from \cite{zhangAdvancedElectricPower2010}: 
\begin{equation}\label{MBO}
\begin{split}
   &\min \limits_{{x},{u}} r_p({x}, {u}, {D},{p},{A}) \\
s.t. \quad& f({x}, {u}, {D},{p},{A}) =0 \\
& \underline{{u}} \leq {u} \leq \bar{{u}} \\
  \quad & \underline{h}_v \leq h_v({x}, {u}, {D},{p},{A}) \leq \bar{h}_v,
\end{split}
\end{equation}
where $r_p$ is the power loss function. ${x}$ is the vector of state variables of the ADN including active power injection, reactive power injection, and voltage magnitude, ${u}$ is the vector of the control variables which are reactive power produced by static var generators (SVGs) and IB-ERs, ${D}$ is the vector of uncontrollable power generations of distributed energy resources and load powers, $p$ denotes parameters of the ADN, ${A}$ is the incidence matrix of the ADN, $f$ is the power flow equation, $\underline{u}$, $\bar{u}$ are the lower and upper bounds of controllable variables,  and $\underline{h}_v$, $\bar{h}_v$ are the lower and upper bounds of voltage.
This paper considers the ADN with $n+1$ buses, and 
bus $0$ is a root bus connected to the main grid.


Recently, the operator usually obtain theoretical parameters of ADNs. Those parameters have some errors while still being reliable to some extent. The inaccurate parameters would degrade the optimization performance of model-based VVC, but it can still be used in real applications \cite{valverdeModelPredictiveControl2013}.

\subsection{Deep Reinforcement Learning based Inverter-based Volt-Var Control}

DRL is a data-driven optimization method that learns the policy to maximize the cumulative reward in the environment. We generally model the problem as a Markov decision process (MDP).
At each step, the DRL agent observes a state ${s}_t$ and generates an action ${a}_t$ according to the policy $\pi$. After executing the action to the environment, the DRL agent observes a reward $r_t$ and the next state ${s}_{t+1}$.
The process generates a trajectory $\tau = \left({s}_{0}, {a}_{0}, r_{1}, {s}_{1}, {a}_{1}, r_{2}, \ldots\right)$.
The infinite-horizon discounted cumulative reward of the RL agent obtained is  $R(\tau)=\sum_{t=0}^{\infty} \gamma^{t} r_{t}$, where $\gamma$ is the discounted factor with $0 \leq \gamma< 1$.
DRL trains the policy $\pi$ to maximize the expected infinite horizon discounted cumulative reward:
\begin{equation}
\pi^{*}=\arg \max _{a \sim \pi} \mathbb{E} [R(\tau)].
\end{equation}

In value-based or actor-critic RL, state-action value function  $Q^{\pi}({s},{a})$ is defined to evaluate the performance of the policy. 
$Q^\pi({s},{a})$ is the expected discounted cumulative reward for starting in state ${s}$, taking an arbitrary action ${a}$, and then acting according to policy $\pi$:
\begin{equation} \label{Q_1}
Q^\pi({s},{a})={\mathbb{E}}_{\tau \sim \pi} \left[\sum_{t=0}^{\infty} \gamma^{t} r_{t} \mid {s}_{0}=s, {a}_{0}=a\right].
\end{equation}


Then the target of DRL is finding optimal policy $\pi^{*}$ to maximize the state-action value function,
\begin{equation}
\pi^{*}=\arg \max _{{a} \sim \pi} Q^\pi({s},{a}).
\end{equation}


For IB-VVC, the inverter-based devices have fast control capability, and the next action is independent of the current action. Then DRL only needs to maximize the immediate reward \cite{sunTwoStageVoltVar2021, caoDataDrivenMultiAgentDeep2021, nguyenThreeStageInverterBasedPeak2022}. Correspondingly, the state action function $Q^\pi({s},{a})$ is:
\begin{equation}\label{one_Q_1}
Q^\pi({s},{a})={\mathbb{E}}\left[ r \mid {s}_{0}={s}, {a}_{0}={a}\right].
\end{equation}


The state $s$, action $a$ and reward $r$ for IB-VVC are defined as follows:
\begin{itemize}
\item[1)] State: ${s} = ({P}, {Q}, {V}, {Q}_{G})$, where ${P}, {Q}, {V}, {Q}_{G}$ are the vector of active, reactive power injection, the voltage of all buses, and the reactive power outputs of controllable IB-ERs and SVGs. Compared with \cite{liuTwoStageDeepReinforcement2021}, adding ${Q}_{G}$ is to reflect the working condition
of the ADN completely. 

\item[2)] Action: 
The action ${a}= {Q}_{G}$, where ${Q}_{G}$ is the reactive power outputs of all IB-ERs and SVGs.
The range of IB-ERs is  $\left|{Q}_{G}\right| \leq \sqrt{{S}_{G}^{2}-\overline{{P}_{G }}^{2}}$, where $\overline{{P}_{G}}$ is the upper limit of active power generation  \cite{yangTwoTimescaleVoltageControl2020,jaakkolaConvergenceStochasticIterative1993}. The range of SVGs is $\underline{{Q}_{G}} \leq {Q}_{G} \leq \overline{{Q}_{G}}$, where $\overline{{Q}_{C}}$ and $\underline{{Q}_{G}}$ is the upper and bottom limit of reactive power generation.
To satisfy the constraints of the controllable variable, the final activation function of actor network is set as ``Tanh". 
Then the output of the final activation function $a_p$ is always in $(-1,1)$. The final action $a_e$ is a linear mapping of $a$ from $(-1,1)$ to the action space. $a_e = 0.5 ( \bar{a} - \underline{a}) a  + 0.5(\bar{a} + \underline{a})$, where $\underline{a}, \bar{a}$ are the upper and bottom bounds. 


\item[3)] Reward: 
The reward for power loss $r_{p}$ is the negative of active power loss 
 \begin{equation}
  r_{p} =  - \sum_{i=0}^n  P_i,
 \end{equation}
and the reward of voltage violation rate $r_v$ is  
\begin{equation}
r_v= -\sum_{i=0}^n \left[\max \left(V_i-\bar{V}, 0\right)+\max \left(\underline{V}-V_i, 0\right)\right].
\end{equation}
The overall reward is 
\begin{equation}
    r = r_p +c_v r_v,
\end{equation} 
where $c_v$ is the penalty factor.
\end{itemize}

\section{Residual Deep Reinforcement Learning with a Reduced Residual Action Space}


To improve the DRL performance in both the initial learning stage and the final stage, this section proposes a residual DRL approach, which learns a residual policy of the base policy from an approximate model-based optimization. 
Firstly, we will introduce the RDRL framework and then combine it with soft actor-critic (SAC) algorithm to design a residual SAC (RSAC).
Considering the IB-VVC has two optimization objectives, we also integrate the two-critic DRL approach into the proposed RSAC.

\subsection{The Framework of Residual Deep Reinforcement Learning}\label{subsection_WMA_SAC}

\begin{figure}[!t] 
\centering
\includegraphics[width=3in]{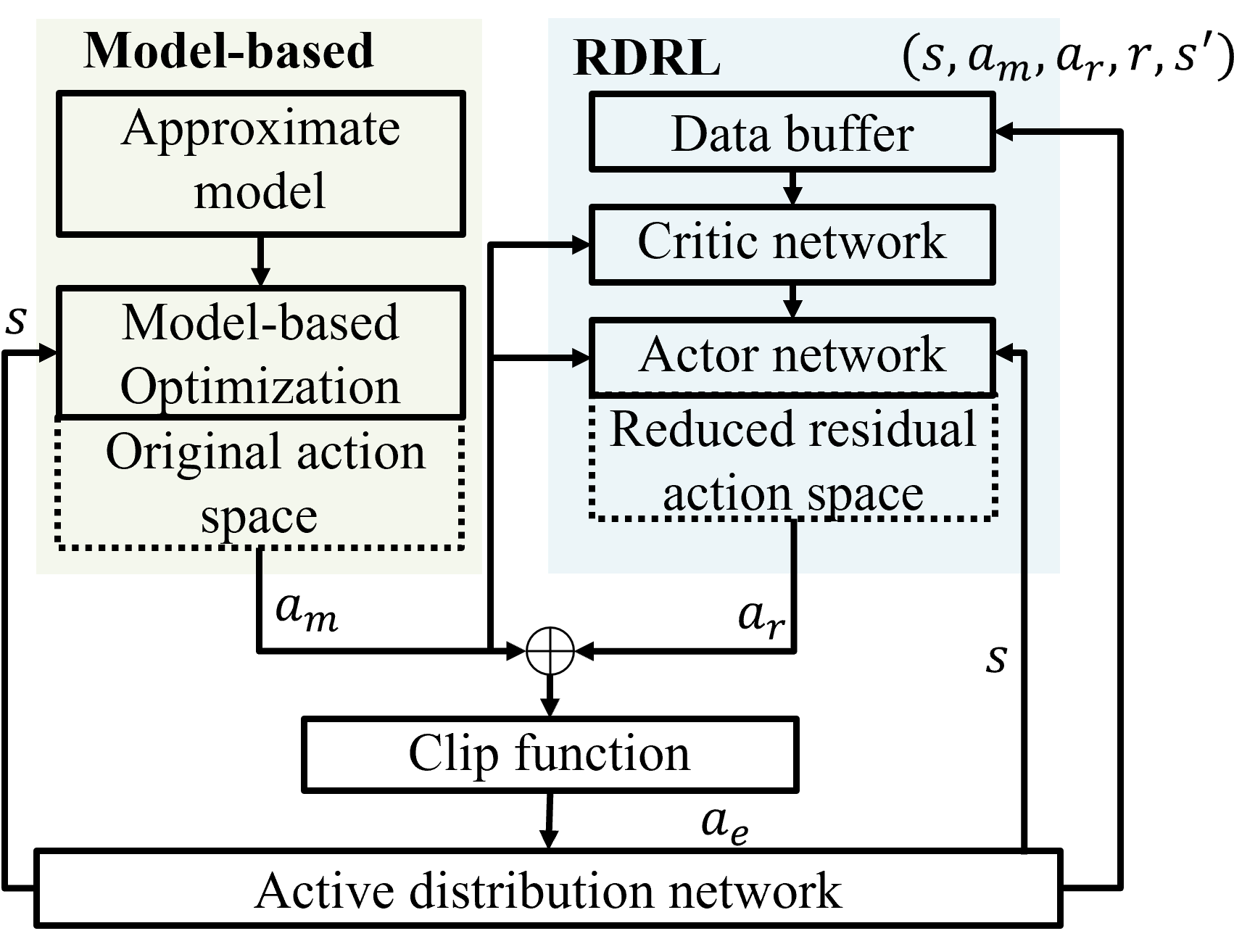}
\caption{The framework of Residual DRL.}
\label{WMA-DRL}
\end{figure}

Fig. \ref{WMA-DRL} shows the framework of the proposed RDRL.
The model-based optimization method calculates the base action ${a}_m$ based on the approximate power flow model in the action space $\mathbb{R}_o  = (\underline{a},\bar{a})$. 
Generally, the model-based approach can acquire a decent VVC performance while there is still a small optimal gap. The difference between the model-based action ${a}_m$  and the optimal action $a$ is named as the optimal residual action $a_r^* = a^*- a_m$. 
Different from general DRL learning the optimal action $ a^*$ directly, 
the RDRL learns the residual action $a_{r}$ of the base action $a_m$.
Since we do not have prior knowledge of $a$, RDRL uses the critic to evaluate the performance of residual action $a_{r}$, and train an actor to output residual action $a_{r}$ that maximizes the critic value.
To further reduce the search difficulties of RDRL, we select a small residual action space $\mathbb{R}_r  = (-\delta,\delta)$. The residual action is set as $\delta = \lambda (\bar{a}-\underline{a})$, where $0<\lambda<1$ is a scale factor.
The final action is $a = a_m + a_r$.

Noting that the final action may be out of the action space of VVC because $\mathbb{R}_o \cup \mathbb{R}_r $ may be large than $\mathbb{R}_o$.
The final action out of the action space needs to be clipped, and thus, the executing action to the ADN is $a_e = \max(\min(a^{k},\bar{a}),\underline{a})$. The clip step is thought of as the internal behavior of ADN environment. Therefore, in DRL algorithms, we still store $a_r$ rather than the clipped value in the data buffer for training. 


The state-action function $Q^{\pi}({s}, {a}_m, {a}_{r})$ for RDRL is:
\begin{equation}\label{Q}
\begin{split}
Q^{\pi_m, \pi_{r}}({s},{a}_m, {a}_{r})=&{\mathbb{E}}_{\tau \sim \pi_m, \pi_{r}} \Big[ \sum_{t=0}^{\infty} \gamma^{t} r_{t} \mid {s}_{0}={s},  \\
&{a}_{m0}={a}_{m}, {a}_{r0}={a}_{r} \Big],
\end{split}
\end{equation}
where $\pi_m$ is the reference-model-based optimization policy and $\pi_{r}$ is the  RDRL policy.


The target of RDRL is to find an optimal policy $\pi_{r}^*$ to maximize the state-action value function,
\begin{equation}\label{gradient_actor}
\pi_{r}^{*}=\arg \max _{a_{r} \sim \pi_{r}} Q^{\pi_m, \pi_{r}}({s}, {a}_m, {a}_{r}).
\end{equation}

The RDRL can be simplified by considering the approximate model-based optimization as an internal behavior of the ADN environment. In this way, $(s,a_m)$ can be viewed as a new state, and DRL can make decisions based on $(s,a_m)$. Additionally, if the model-based optimization solver is deterministic that one state ${s}$ corresponding to only one action ${a}_m$, then $Q^{\pi_m, \pi_{r}}({s},{a}_{m}, {a}_{r}) = Q^{\pi_{r}}({s},{a}_{r})$ and $a_m$ can be omitted in the new state $(s,a_m)$.

RDRL has three key rationales to improve VVC performance:


\textbf{Inheriting the capability of the model-based optimization with an approximate model:}
The model-based optimization approach with an approximate model has a decent VVC performance. The final action is the superposition of the action of model-based optimization and the output of the actor of RDRL.
In the initial learning stage, the actor of RDRL has no optimization capability, and the output is close to zero. The model-based optimization approach mainly provides the VVC capability.


\textbf{Residual policy learning:}
Similar to the boosting regression in supervised learning \cite{friedmanGreedyFunctionApproximation2001}, the actor learns the residual action between the global optimal action and the action of the Oracle model-based optimization. It reduces the learning difficulties of the actor and enhances the optimization performance of RDRL. 

\textbf{Learning in a reduced action space:}
The advantages of learning in reduced action space can be divided into twofold. 
\begin{itemize}
    \item[1.] Reduced action space leads to a smaller approximation error of the critic. The critic approximate the state-action function, when we reduce the space of action, the approximation space decreased. Neural network approximate in a smaller space would be more accurate. 
    It provides a more accurate gradient for the training of the actor.
    \item[2.] Reduced action space reduces the exploration difficulties of actor. Intuitively, it would be easier for the actor to search for the optimal residual action in a smaller residual action space.
    
\end{itemize}
We will verify the benefits of each rationale through simulation.

\subsection{Residual Soft Actor-Critic}\label{RSPTCSAC}

\begin{algorithm}[!t]
  \caption{Residual Soft Actor-Critic} \label{reference-model-assited-algorihtm}
  \begin{algorithmic}[1]
  \Require
  Initial policy parameters ${\theta}$, Q-function parameters ${\phi}_1$, ${\phi}_2$, and replay buffer $\mathcal{D}$.
  \Require approximate power flow model.
  \Require Set the scale factor $\lambda$ for residual action space.
\For {t = 1 \textbf{to} T}
\State Given $s$, calculate the model-based action ${a}_m$. 
\State Calculate the residual action ${a}_{r} \sim \pi_{r}^{{\theta}}(\cdot \mid {s, a_m})$.

\State  The action $a = {a}_m + {a}_r$, the execution action is 
\Statex \quad \  $a_e = \min(\max(a,\underline{a}),\bar{a})$.
\State Store $({s},a_m, a_{r},r_p,r_v, s^{\prime})$ in replay buffer $\mathcal{D}$.
\If { $t> t_1$}
\For {$j$ in range (how many updates)} 
 \State Randomly sample a batch of transitions
\Statex \qquad \qquad \ $B={(s,a_m, a_r,r_p,r_v, s^\prime )}$ from $\mathcal{D}$.
\State Update $Q_{{\phi}_p}$ and $Q_{{\phi}_v}$ by minimizing  \eqref{L_one_step_Q}
\State Update $\alpha$ by minimizing \eqref{entropy_adjust}.
\State Update $\pi_{rp}^{{\theta}}$ by maximizing \eqref{loss_actor}.
\EndFor
\EndIf
\EndFor
\end{algorithmic}
\end{algorithm}

The RDRL approach is compatible with most actor-critic DRL algorithms like DDPG \cite{lillicrapContinuousControlDeep2019}, TD3 \cite{fujimotoAddressingFunctionApproximation2018}, and SAC \cite{haarnojaSoftActorCriticOffPolicy2018}. 
Here, we select SAC as the baseline and propose a residual SAC (RSAC). 
SAC has the following four critical tricks to improve its learning performance: 1) replay buffers, 2) target networks, 3) clipped double-Q Learning, 4) entropy regularization. The former two tricks are inherited from DDPG to improve the learning stability \cite{lillicrapContinuousControlDeep2019}.  The third trick is inherited from TD3 to address the overestimation of critic network \cite{fujimotoAddressingFunctionApproximation2018}. SAC also proposes entropy regularization to achieve a more stable Q-value estimation and improve exploration efficiency \cite{haarnojaSoftActorCriticOffPolicy2018}.

Unlike the SAC learning a policy directly, RDRL learns a residual policy of the model-based optimization method   $\pi_m$ under an approximate model.
RDRL maximizes the entropy-regularized discounted accumulated reward by optimizing the residual policy,
\begin{equation}\label{actor}
\pi_r^*=\arg \max _{\pi_r} \mathbb{E}_{\tau \sim \pi_m, \pi_r}\left[\sum_{t=0}^{\infty} \gamma^t\left(r_t+\alpha H\left(\pi\left(\cdot \mid s_t\right)\right)\right)\right].
\end{equation}

The entropy-regularized critic in RSAC is
\begin{equation}
\begin{split}
 Q^{\pi_{m}, \pi_{r}}({s}, a_m, {a}_{r})= &{\mathbb{E}}_{\tau \sim \pi_{m}, \pi_{r}} \Big[\sum_{t=0}^{\infty} \gamma^{t} r_{t}+\alpha \sum_{t=1}^{\infty} \gamma^{t} H\left(\pi\left(\cdot \mid {s}_{t}\right)\right) \\ 
 & \mid {s}_{0}={s}, a_{m0} = a_m, 
  {a}_{r0}={a}_{r} \Big], 
 \end{split}
\end{equation}
where $H\left(\pi_{rp}\left(\cdot \mid {s}_{t}\right)\right)=\underset{{a} \sim \pi\left(\cdot \mid {s}_{t}\right)}{\mathbb{E}}\left[-\log \pi_{rp}\left(\cdot \mid {s}_{t}\right)\right]$ is the entropy of the stochastic policy at ${s}_{t}$, $\alpha$ is the temperature parameter.

RDRL optimizes the residual policy $\pi_{r}^*$ to maximize the entropy-regularized critic,
\begin{equation}\label{gradient_actor}
\pi_{r}^{*}=\arg \max_{a_{r} \sim \pi_{r}} Q^{\pi_m, \pi_{r}}({s}, {a}_m, {a}_{r}) +  \alpha H\left(\pi\left(\cdot \mid {s}_{t}\right)\right).
\end{equation}

For IB-VVC tasks, DRL only needs to maximize the immediate reward rather  than the long-horizontal accumulated reward \cite{sunTwoStageVoltVar2021, caoDataDrivenMultiAgentDeep2021, nguyenThreeStageInverterBasedPeak2022}. It reduces the learning difficulties of the DRL task and avoids the overestimation of the critic network.
The critic of RSAC for IB-VVC can be simplified as
\begin{equation}\label{one_Q}
Q^{\pi_m, \pi_{r}}(s,{a}_{m}, {a}_{rp})={\mathbb{E}}\left[ r(s,{a}_m, {a}_{r}) \right].
\end{equation}

Considering two objectives of IB-VVC: minimizing power loss and eliminating voltage violations have different mathematical properties, we can integrate two-critic DRL into the RSAC algorithm to improve learning speed and optimization capability \cite{liuReducingLearningDifficulties2022}. The two-critic DRL approach utilizes two critics to approximate the rewards of two objectives separately, which reduces the learning difficulties of each critic. 
The critics of minimizing power loss and eliminating voltage violations $Q_p, Q_v$ are 
\begin{equation}\label{two_Qp}
\begin{split}
Q_p^{\pi_m, \pi_{r}}(s,{a}_{m}, {a}_{r}) &={\mathbb{E}}\left[ r_p(s,{a}_m, {a}_{r}) \right],\\
Q_v^{\pi_m, \pi_{r}}(s,{a}_{m}, {a}_{r}) &={\mathbb{E}}\left[ r_v(s,{a}_m, {a}_{r}) \right].
\end{split}
\end{equation}
 


In real applications, SPTC-RDRL stores the historical data $(s,a_m,a_r,s^\prime)$ into a data buffer $\mathcal{D}$, and then samples mini-batch data $\mathcal{B}$ from the data buffer to train both the actor and critic neural networks in each training step.
The critic networks of power loss and voltage violation are learned by minimizing the loss function 
\begin{equation} \label{L_one_step_Q}
\begin{split}
     L_{{\phi}_{p}}&=\frac{1}{|\mathcal{B}|}\sum_{ \mathcal{B} \sim \mathcal{D}}\left[\left(Q_{{\phi}_{p}}({s}, a_m, {a}_{r})-r_p\right)^{2}\right],\\  
     L_{{\phi}_{v}}&=\frac{1}{|\mathcal{B}|}\sum_{ \mathcal{B} \sim \mathcal{D}}\left[\left(Q_{{\phi}_{v}}({s}, a_m, {a}_{r})-r_v\right)^{2}\right],
\end{split}
\end{equation}
where $|\mathcal{B}|$ is the number of the mini-batch data, $\phi_p, \phi_v$ are the parameters of critic networks of power loss and voltage violations.


Similar to SAC, the residual actor $\pi_{r}^{{\theta}}$ is a stochastic policy which is parameterized as
\begin{equation}
\begin{split}
\pi_{r}^{{\theta}}(\cdot \mid s, a_m)= & \tanh \left(\mu_{{\theta}}(s,a_m)+\sigma_{{\theta}}(s,a_m) \odot \xi\right), \\
& \quad \xi \sim \mathcal{N}(0, I),
\end{split}
\end{equation}
where ${\theta}$ is the parameters of actor network, $\mu$ is the mean function, and $\sigma$ is the variance function. $\pi_{r}^{{\theta}}$ is learned by maximizing the loss function $L_{{\theta}}$,
\begin{equation}\label{loss_actor}
\begin{split}
  L_{{\theta}} =&  \frac{1}{|B|}\sum_{\mathcal{B}| \sim \mathcal{D}} \Big[ Q_{{\phi}_{p}}\left({s},a_m, \pi_r^{\theta}(s,a_m)\right) \\
  & +  Q_{{\phi}_{v}}\left({s},a_m, \pi_r^{\theta}(s,a_m)\right) -\alpha \log \pi_{r}^{{\theta}}\left( \cdot \mid s, a_m \right)\Big].
\end{split}
\end{equation}

The entropy regularization coefficient $\alpha$ can be a constant or adjustable dual variable by minimizing the loss function $ L(\alpha)$,
\begin{equation}\label{entropy_adjust}
    L(\alpha) =  \frac{1}{|B|} \sum_{\mathcal{B} \in B } [-\alpha \log \pi_{{\theta}} ({\cdot}|s,a_m) - \alpha \mathcal{H}],
\end{equation}
where $\mathcal{H}$ is the entropy target. 

In training stage, the residual action ${a}_r$ is sampling from the stochastic policy $\pi_{r}^{{\theta}}$. In the testing stage, $\sigma_{\theta}$ is settled as 0, and  the policy $\pi_{r}^{{\theta}}$ works in its deterministic mode. 
After superposing the action of model-based optimization action, the final action is ${a} = {a}_m +  {a}_r$, and the executing action to the ADN is $a_e = \min(\max(a,\underline{a}),\bar{a})$.


RSAC is shown in Algorithm \ref{reference-model-assited-algorihtm}. In training the RSAC agent, we propose the following tricks to achieve a better learning performance:
\begin{itemize}
\item[1.] In the initial learning stage, the actor has no VVC capability.
The action $a_m$ of model-based optimization should dominate the control process, and the actor should output a value near zeros. To achieve this, we initialize the weights of the final layer of the actor with a uniform distribution in the interval $[-0.001,0.001]$. We cannot initialize all weights of the neural networks as zeros because it is difficult to train.  

\item[2.] DRL learns from a large scale of data, while in the initial learning stage, the amount of data is small. Neural networks constantly learning from a small number of data may be overfitting. It would affect the subsequent learning process. 
To alleviate it, RDRL begins to learn after generating more data. Therefore, the step to start learning $t_1$ is set, which is generally $5-20$ times the batch size.


\end{itemize}

\subsection{Discussion}

Our work is closely related to the existing residual reinforcement learning \cite{silverResidualPolicyLearning2018, johanninkResidualReinforcementLearning2019,zhangResidualPolicyLearning2022}, but has some distinct differences:
\begin{itemize}
    \item[1.] Existing RDRL overlook the importance of a reduced residual action space in improving optimal results. In some cases, the residual action space may even be set at twice the size of the original action space to ensure full coverage of the original action space \cite{zhangResidualPolicyLearning2022}. However, an excessively large action space would deteriorate learning performance. For the problem that residual action space cannot cover the optimal action, we propose a boosting DRL in section \ref{BRDRL_s}.
    \item[2.] Existing RDRL focuses on using the base policy to guide exploration in RDRL, while this paper focuses on exploitation that reduces the approximation error of critic and the learning difficulties of actor. The reason for this may be that the studied tasks are completely different. Concurrent works on residual reinforcement learning study the long-horizon, sparse-reward problem  \cite{silverResidualPolicyLearning2018} or very difficult learning problems like contact-intensive tasks \cite{johanninkResidualReinforcementLearning2019}. 
    Achieving optimal results with general DRL algorithms can be challenging, as they may struggle to converge and become unstable. 
    In this paper, IB-VVC is a reward-intensive problem. DRL on IB-VVC is easier to converge, and the learning process is stable.
    The optimal gap is very small after sufficient time to learn. After applying RDRL with a reduced action space in IB-VVC, the optimal gap declines further.
\item[3.] Existing RDRL only demonstrates the superiority of RDRL in the whole training process in the simulation. However, they neither mention  ``residual policy learning" and ``learning in a reduced action space" nor verify them in simulation. This paper designs simulations that verify the three key rationales of RDRL point-by-point. 
\end{itemize}

\section{Boosting Residual Deep Reinforcement Learning}\label{BRDRL_s}

\begin{figure*}[!t]
\centering 
\subfigure[``too small" residual action space]{\begin{minipage}[t]{0.25\linewidth}\label{too_small}
    \includegraphics[width=1.4in]{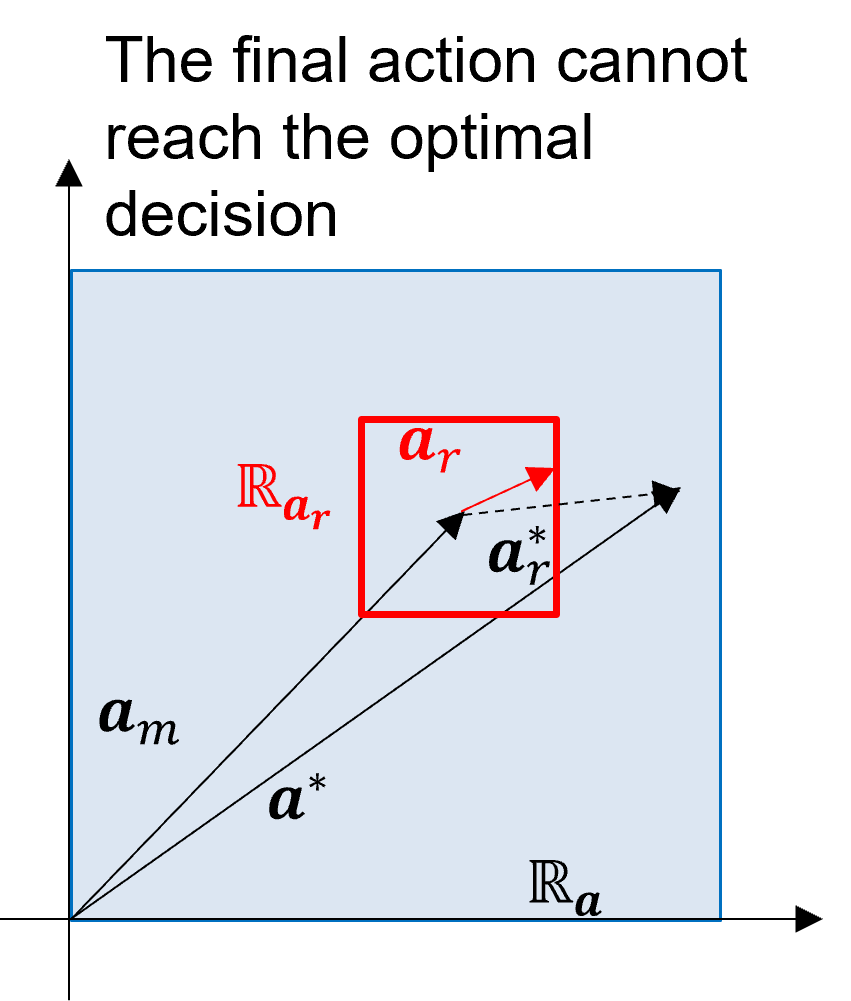}
    \end{minipage}
    }
\subfigure[``too large" residual action space]{
\begin{minipage}[t]{0.25\linewidth}\label{too_large}
\includegraphics[width=1.4in]{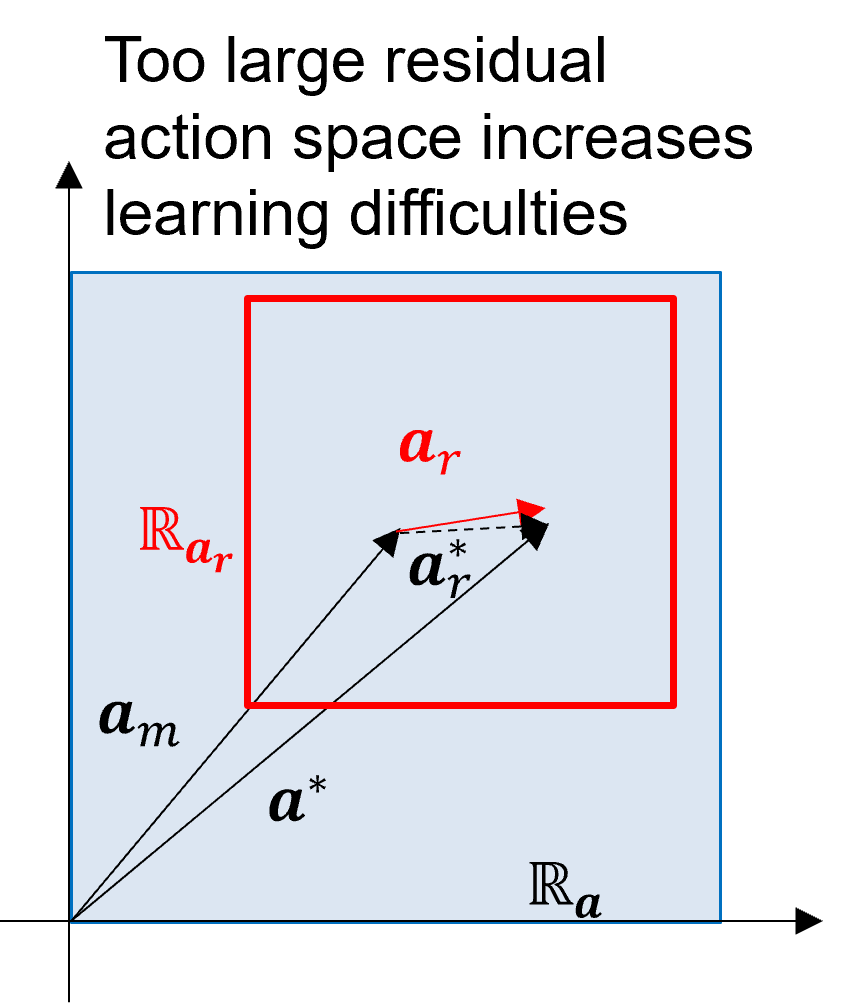}
\end{minipage}
    }
\subfigure[Iterating DRL with a small residual action space]{
\begin{minipage}[t]{0.25\linewidth}\label{iterating_several_time}
\includegraphics[width=1.4in]{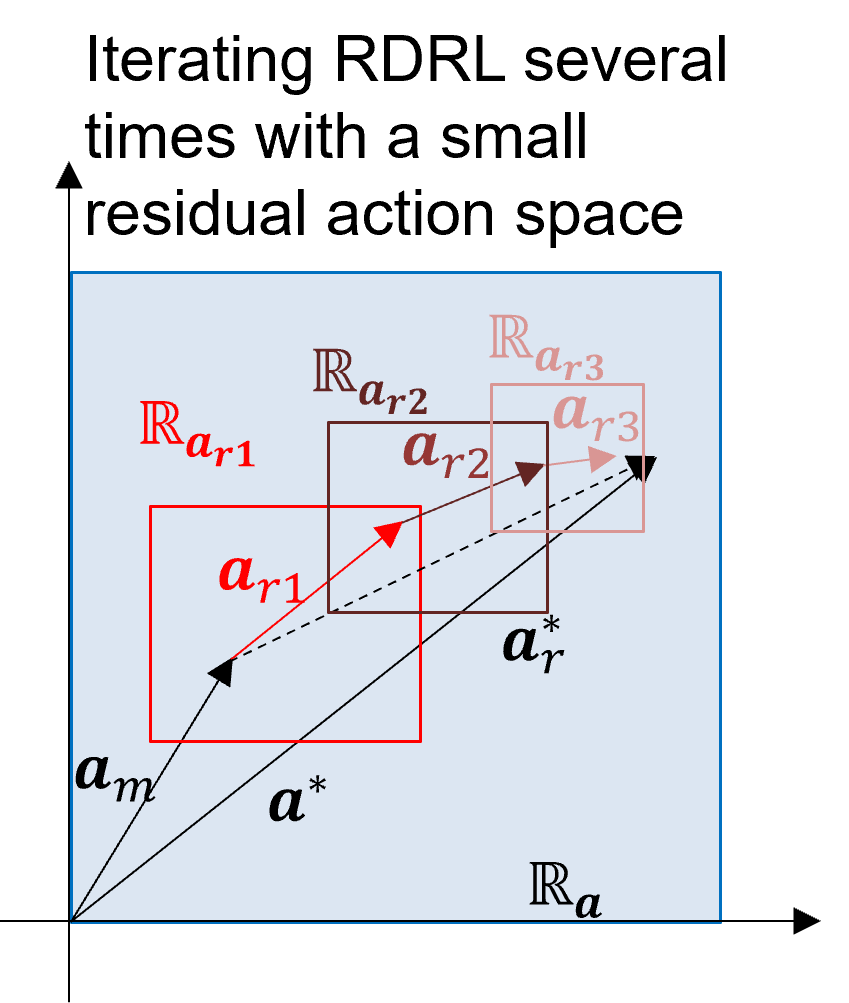}
\end{minipage}
}
\caption{The problem of ``too small" or ``too large"  residual action space and their solution: boosting residual DRL (BRDRL).}\label{too_problem}
\end{figure*}

Learning in a reduced residual action space is one of the key reasons to improve the optimization capability of RDRL. However, it is challenging to determine the residual action space because the optimal action and residual action are unknown beforehand. As shown in Figs. \ref{too_small}, \ref{too_large}, if the residual action space is  "too small", it would restrict the residual action reach to the optimal action, while if the residual action space is "too large", it hinders RDRL perform the full potential because of learning in a large action space.

To further improve the optimization performance of RDRL, we propose a boosting RDRL (BRDRL) to train a sequential RDRL, as shown in Fig. \ref{iterating_several_time}.
In each learning process, the base policy is from the last training, and the RDRL learns a residual action in a further reduced residual action space. The final action would reach close to the optimal action in each learning process.
It is similar to the boosting regression in supervision learning, while the difference is that BRDRL needs to reduce the residual action space for each RDRL.

The RDRL in section \ref{RSPTCSAC} is seen as the first RDRL,
\begin{equation}
\begin{split}
    {a}^{1}  &= {a}_{m} + {a}_{r}^{1}\\
    {a}_{e}^{1} &= \max(\min(a^{1},\bar{a}),\underline{a}).
\end{split}
\end{equation}

In the $k^{th}$ RDRL, we first select a reduced residual action space $(-\delta^k, \delta^k)$, where $\delta^k = \lambda_k (\bar{a} - \underline{a})$, and then train RDRL. The final action is
\begin{equation}\label{boosting_action}
\begin{split}
{a}^{k} &=  {a}_{e}^{k-1} + {a}_{r}^{k}\\
{a}_{e}^{k} &= \max(\min(a^{k},\bar{a}),\underline{a}).
\end{split}
\end{equation}


The data buffer stores $(s, a^{k-1}, a^{k}, r, s^\prime)$ in each time step. $s^\prime$ is the next state. 
We see the clip function ${a}_{e}^{k} = \max(\min(a^{k},\bar{a}),\underline{a})$ as the internal behavior of the environment.
In the train $k^{th}$ RDRL, $ k^{th}$ actor is stochastic policy, the actor of $1, \dots k-1^{th}$ RDRL should work in deterministic mode that set $\sigma_{\theta} = 0$ in equation \eqref{actor}, and the corresponding residual action is $a_r^{i} = \mu_{\theta_i}(s,a^{i-1})$. 

The boosting RDRL can alleviate the problems of ``too small"  or `too large" residual action space mentioned in  section \ref{subsection_WMA_SAC}. 
For the ``too small"  residual action space, in the next learning process, we can set a small residual action space. After the next RDRL, the residual action would be close to the optimal action step by step. 
For the ``too large" residual action space, we can also set a small residual action space in the next iteration, it can improve the optimization performance by reducing the residual action space. 

Generally, the parameter error of the approximate model is not large, and the decision is close to the optimal decision.
In addition, with the increasing number of iterations, the decision would be closer to the optimal decision, and the improved rate of the iterating RDRL decreases. 
Therefore, boosting the approach twice or thrice is enough in a real application.
In the first RDRL, we generally set the residual action space as $0.4 - 0.6$ times the original action space, and in the second, set $0.1-0.3$ times. 

The boosting RSAC (BRSAC) can be obtained by a few modifications of RSAC in section \ref{RSPTCSAC}. In training $k^{th}$ RDRL, we input the approximate power flow model and the actors of the $1^{th} \dots (k-1)^{th}$ RDRL.
In step 2, we calculate the model-based action $a_m$, the $1^{th} \dots k-1^{th}$ residual actions and then obtain $a^{k-1}, a_e^{k-1}$ according to \eqref{boosting_action}.
In step 4, the action is calculated by $a^k = a_e^{k-1} + a_r^k$.

\section{Simulation}

Numerical simulations were conducted on 33-bus, 69-bus, and 118-bus distribution networks downloaded from MatPower \cite{zimmermanMATPOWERSteadyStateOperations2011}. 
In the 33-bus system, 3 IB-ERs were connected to bus 17, 21, and 24, and 1 SVG of 2 MVar was connected to bus 32.   
In the 69-bus system, 4 IB-ERs were connected to bus 5, 22, 44, and 63, and 1 SVG was connected to bus 13.
In the 118-bus system, 8 IB-ERs were connected to bus 33, 50, 53, 68, 74, 97, 107, and 111,  and 2 SVGs were connected to bus 44 and 104.
Each IB-ER had 1.5 MW active power and 2 MVar reactive power capacity. Each SVG had 2 MVar reactive power capacity.
All load and generation levels were multiplied with the fluctuation ratio \cite{liuTwoStageDeepReinforcement2021} and a $20\%$ uniform distribution noise to reflect the variance.
The algorithms were implemented in Python. The balanced power flow was solved by Pandapower \cite{thurnerPandapowerOpenSourcePython2018} to simulate ADNs, and the implementation of the DRL algorithms is based on PyTorch.
The base of powers is set as 1 MVA.

The following methods were compared in our simulations:
\begin{itemize}
    \item[1)] \textbf{Model-based optimization under an accurate model (MBO):}
    Model-based optimization was solved by PandaPower with the interior point solver.
    We assumed the original parameters in test distribution networks were accurate. Generally, accurate parameters are difficult to obtain. 
    The result of model-based optimization with an accurate power flow model was an ideal result and is taken as a baseline for evaluating the performance of DRL algorithms.
    \item[2)] \textbf{Model-based optimization under an approximate model (AMBO):} In the approximate model, the approximate parameters of resistance and reactance of branches were set as 1.5 times the original ones for 33-bus and 69-bus distribution networks, and 1.3 times for 118-bus distribution network. We solved the VVC tasks on the approximate model and test it on the accurate model.
    \item[3)] \textbf{DRL:} Considering IB-VVC has two optimization objectives, we used the two critic SAC adopted from the paper \cite{liuReducingLearningDifficulties2022}.
    \item[4)] \textbf{RDRL:} The RDRL proposed in section \ref{RSPTCSAC} is also developed based the two critic SAC approach. RDRL combined model-based optimization under an approximate model and DRL.
    The simulation contained 10 experiments. For each experiment, ${\delta} = \lambda *  (\bar{{a}}-\underline{{a}})/2$, where the scale factor $\lambda = 0.1,0.2,\dots 1$. 
    \item[5)] \textbf{BRDRL:} The proposed BRDRL trained upon the results of 10 RDRL experiments. So it also contained 10 experiments. ${\delta} = 0.2 * (\bar{{a}}-\underline{{a}})/2$ for the residual action space.
    The simulation can show that BRDRL can alleviate the issues of ``too small" or ``too large" residual action spaces of RDRL.
\end{itemize}

We trained the DRL agent using 300 days of data. The step to start learning $t_1$ is 960 for both DRL, RDRL, and BRDRL.
The other hyper-parameters of the three DRL algorithms
 were the same as in our previous paper \cite{liuReducingLearningDifficulties2022}.
We tested the DRL algorithms in the same environment at each step in the training process. We stored the training results and testing results at each step.

In this section, we first verified the superiority of RDRL and BRDRL regarding reward, power loss, and voltage violation. Then, we verified the reasons for the superiority of RDRL point by point. 
Finally, we verified that BRDRL can improve the optimization performance further. 

\subsection{The Superiority of RDRL and BRDRL}

\begin{figure*}[ht]
\centering
\subfigure[33-bus]{\begin{minipage}[t]{0.315\linewidth}\label{wm_DRL_result_33}
    \includegraphics[width=2.4in]{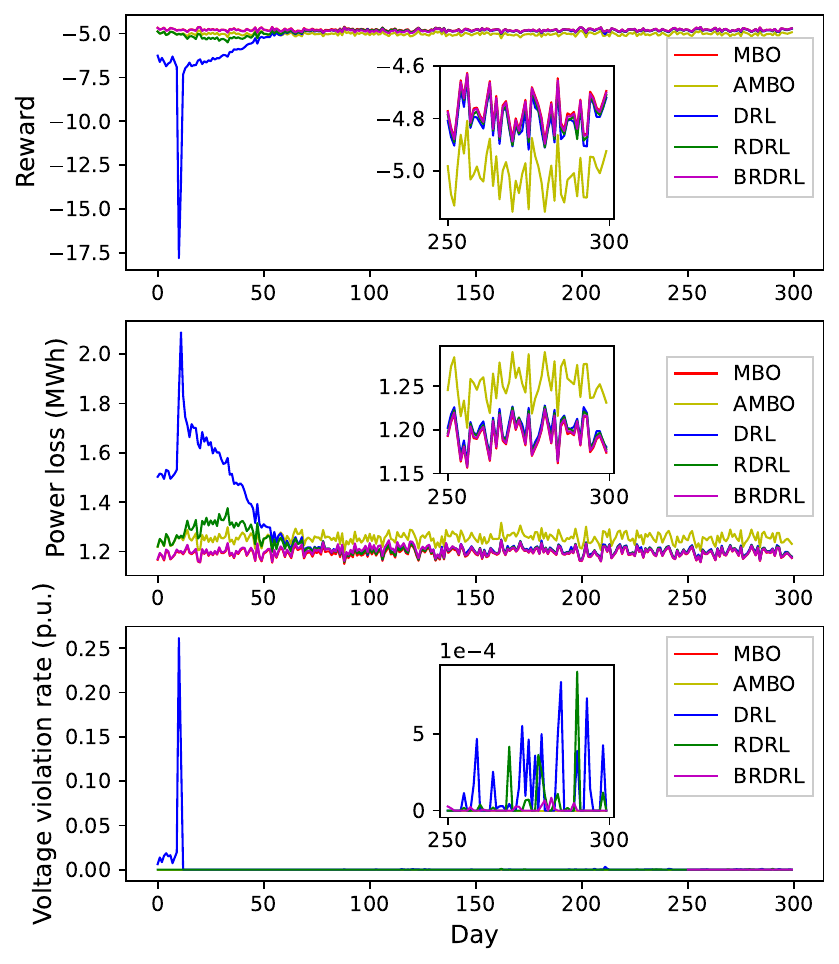}
    \end{minipage}
    }
\subfigure[69-bus]{
\begin{minipage}[t]{0.315\linewidth}\label{wm_DRL_result_69}
\includegraphics[width=2.4in]{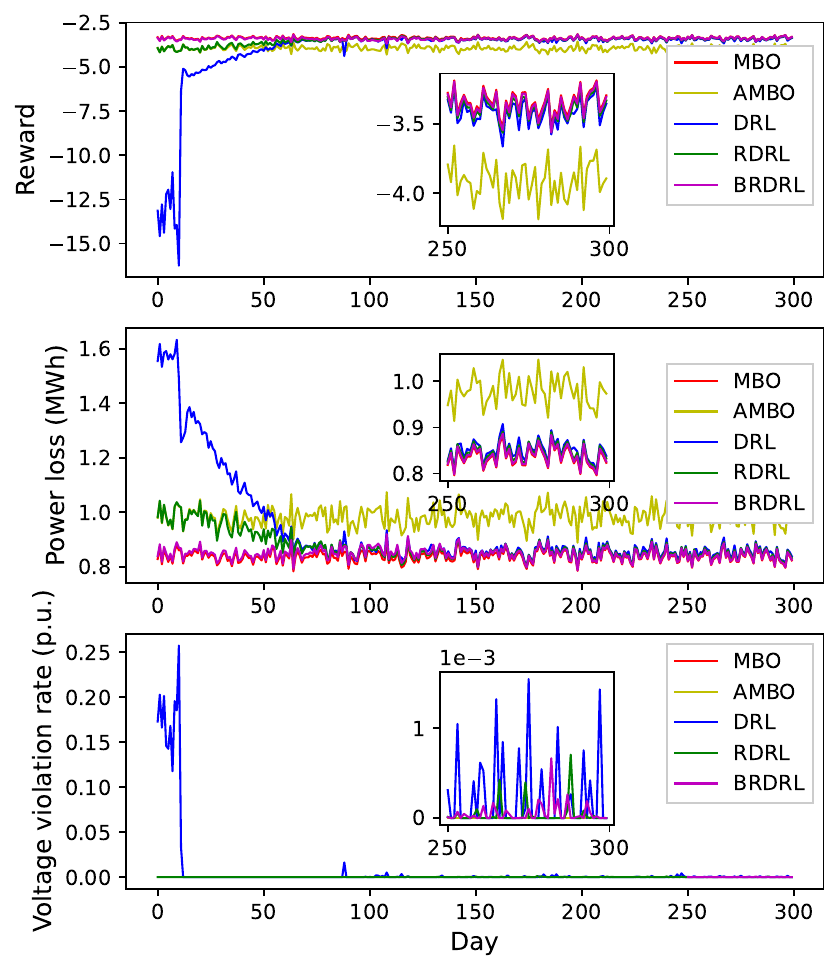}
\end{minipage}
    }
\subfigure[118-bus]{
\begin{minipage}[t]{0.315\linewidth}\label{wm_DRL_result_118}
\includegraphics[width=2.4in]{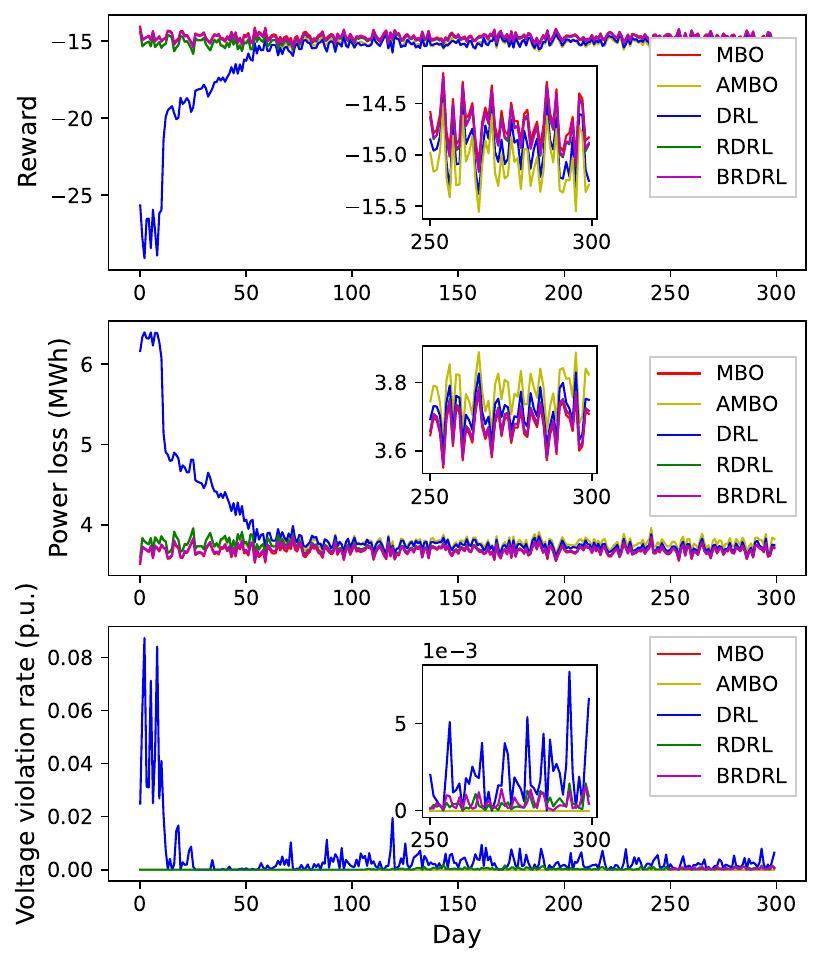}
\end{minipage}
    }
\caption{The testing results of the model-based optimization with an accurate model (MBO), the model-based optimization with an approximate model (AMBO),
deep reinforcement learning (DRL), residual DRL (RDRL), and boosting RDRL (BRDRL) }\label{wm_DRL_result}
\end{figure*}

\begin{figure*}[ht!]
\centering
\subfigure[33-bus]{\begin{minipage}[t]{0.315\linewidth}\label{error_result_33_all}
    \includegraphics[width=2.4in]{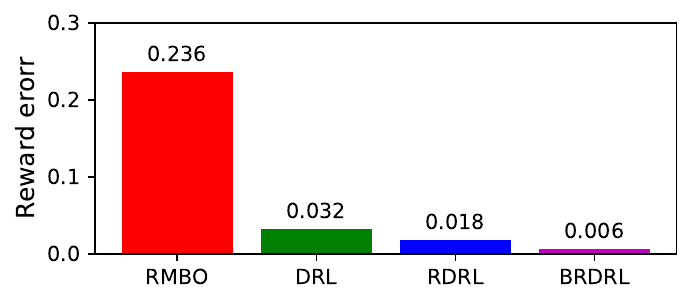}
    \end{minipage}
    }
\subfigure[69-bus]{
\begin{minipage}[t]{0.315\linewidth}\label{error_result_69_all}
\includegraphics[width=2.4in]{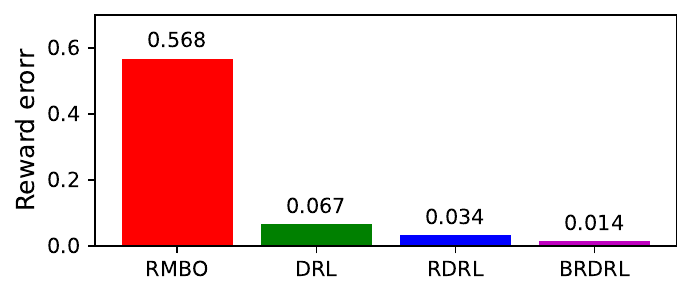}
\end{minipage}
    }
\subfigure[118-bus]{
\begin{minipage}[t]{0.315\linewidth}\label{error_result_118_all}
\includegraphics[width=2.4in]{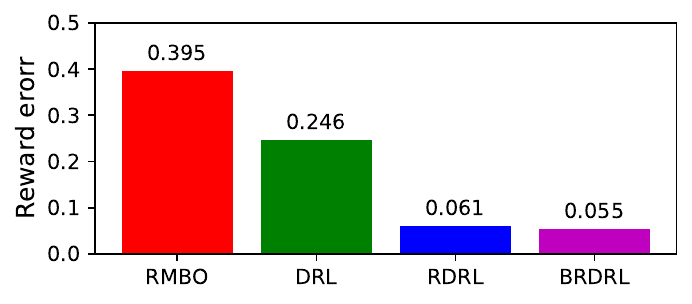}
\end{minipage}
    }
\caption{The reward results of the model-based optimization with an accurate model (MBO), the model-based optimization with an approximate model (AMBO),
deep reinforcement learning (DRL), residual DRL (RDRL), and boosting RDRL (BRDRL) in the final 50 episodes. 
Here, the reward error = the result of model-based optimization with an accurate model - the result of the mentioned method.}
\label{error_result}
\end{figure*}


The superiority of RDRL and BRDRL can be verified by comparing the results of DRL and AMBO in Fig. \ref{wm_DRL_result}.
We plot the result $\lambda = 0.5$ for RDRL, and $\lambda_1 = 0.5, \lambda_2 = 0.2$ for BRDRL.

We made two observations. 

First and foremost, in the initial learning stage, RDRL and BRDRL performed better than SAC regarding reward, power loss, and violation rate. In days 1-10, RDRL inherited the VVC capability of AMBO, the output of the actor of RDRL is initialized to close to zeros, so it has a similar performance as AMBO. After day 10, the RDRL began to learn. The performance had a small fluctuation and then converged to a higher reward.
Similarly, BRDRL inherited the VVC capability of RDRL, so it had a better VVC performance than AMBO, DRL, and RDRL in the initial learning stage. Since the residual action space is 0.2 times the original action space, the fluctuation is even invisible.




Second, in the final learning stage (days 250 to 300), RDRL and BRDRL had a considerably better VVC performance than AMBO and slightly better than DRL regarding reward, power loss, and violation rate.
The super performance is because RDRL and BRDRL are learned in the residual action with a reduced residual action space.
AMBO is the worst because it utilizes the approximation model.


Fig. \ref{error_result} shows the quantified optimization results error of the four methods. We saw the results of the model-based optimization method with an accurate model as the optimal results. 
Compared to DRL, RSAC reduced the reward error to $44\%, 50\%, 75\%$ in 33, 69, and 118 bus distribution networks respectively. 
The VVC performance is improved further by BRDRL. Compared to SAC, BRSAC reduced the reward error to $81\%, 80\%, 78\%$ in 33, 69, and 118 bus distribution networks, respectively.

\subsection{The Rationales of A Residual Deep Reinforcement Learning}

As we discuss on \ref{subsection_WMA_SAC}, three key components improve the RDRL performance: 1)inherited the capability of the model-based optimization, 2) residual policy learning, and 3) learning in a reduced residual action space. 
The advantage of ``inherited the capability of the model-based optimization" had been verified in Fig. \ref{wm_DRL_result} by comparing RDRL and SAC in the initial learning stage. This subsection would verify the latter two rationales. 

\begin{figure*}[ht]
\centering
\subfigure[33-bus]{\begin{minipage}[t]{0.315\linewidth}\label{wm1_large_sacle_33}
    \includegraphics[width=2.4in]{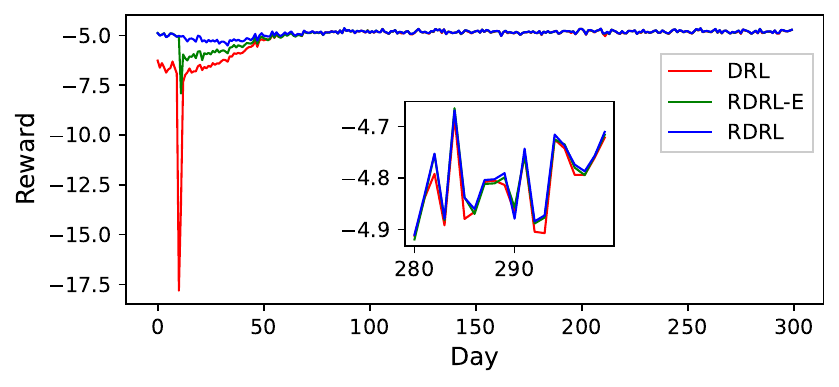}
    \end{minipage}
    }
\subfigure[69-bus]{
\begin{minipage}[t]{0.315\linewidth}\label{wm3_large_sacle_69}
\includegraphics[width=2.4in]{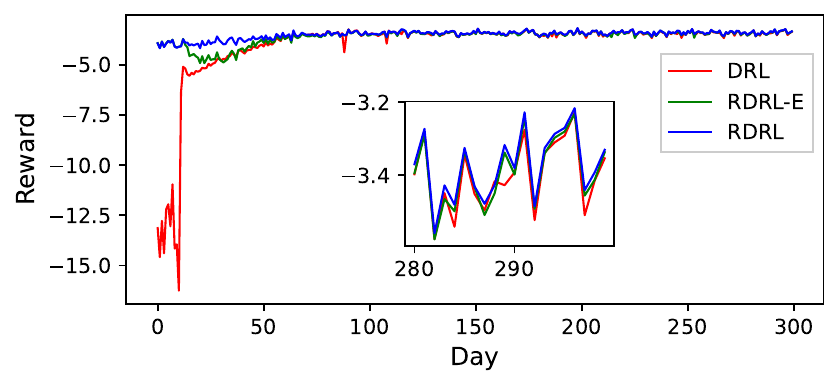}
\end{minipage}
    }
\subfigure[118-bus]{
\begin{minipage}[t]{0.315\linewidth}\label{wm3_large_sacle_118}
\includegraphics[width=2.4in]{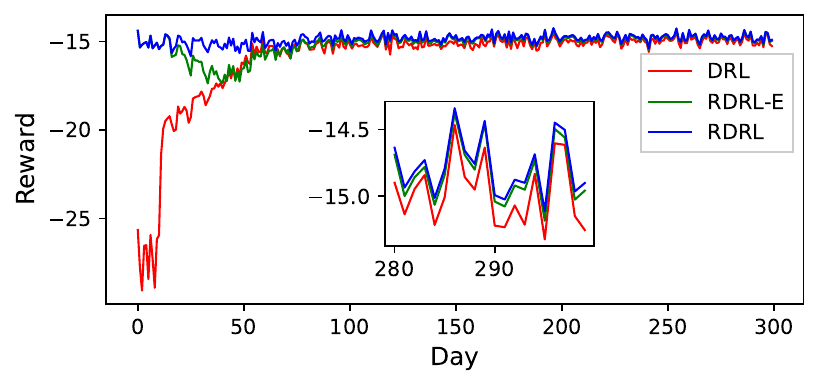}
\end{minipage}
    }
\caption{The reward error of three DRL algorithms: original DRL, residual DRL with an equal residual action space as the original action space (RDRL-E), and residual DRL with a reduced residual action space (RDRL). The scale factor of residual action space $\lambda = 1$ for RDRL-E, and $\lambda = 0.5$ for RDRL. 
} \label{RDRL_E_trajectory}
\end{figure*}

\begin{figure*}[ht]
\centering
\subfigure[33-bus]{\begin{minipage}[t]{0.315\linewidth}\label{wm1_large_sacle_33}
    \includegraphics[width=2.4in]{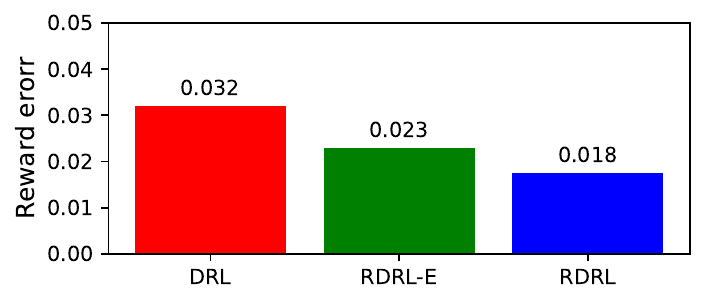}
    \end{minipage}
    }
\subfigure[69-bus]{
\begin{minipage}[t]{0.315\linewidth}\label{wm3_large_sacle_69}
\includegraphics[width=2.4in]{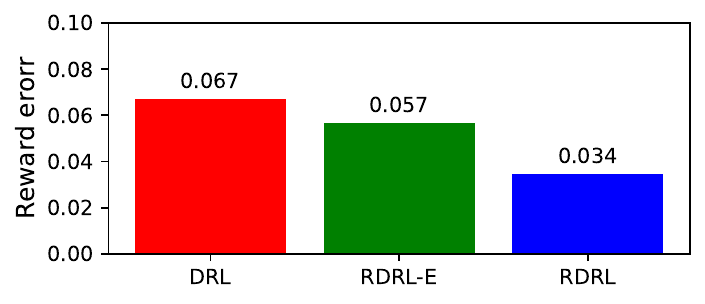}
\end{minipage}
    }
\subfigure[118-bus]{
\begin{minipage}[t]{0.315\linewidth}\label{wm3_large_sacle_118}
\includegraphics[width=2.4in]{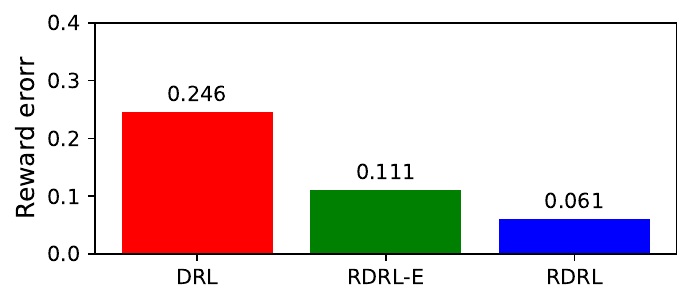}
\end{minipage}
    }
\caption{The reward error of DRL, Residual DRL with equal residual action space as the original action space (RDRL-E), Residual DRL with a reduced residual action space in final 50 episodes. $\lambda = 1$ for RDRL-E, and $\lambda = 0.5$ RDRL. 
} \label{RDRL_E}
\end{figure*}

\begin{figure*}[ht]
\centering
\subfigure[33-bus]{\begin{minipage}[t]{0.315\linewidth}\label{wm1_large_sacle_33}
    \includegraphics[width=2.4in]{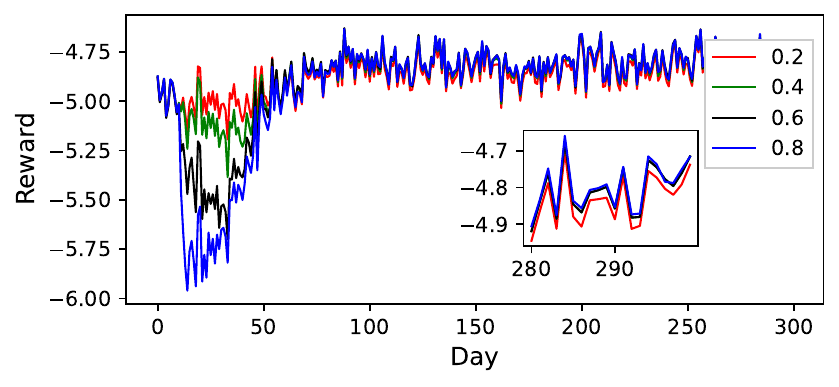}
    \end{minipage}
    }
\subfigure[69-bus]{
\begin{minipage}[t]{0.315\linewidth}\label{wm1_large_sacle_69}
\includegraphics[width=2.4in]{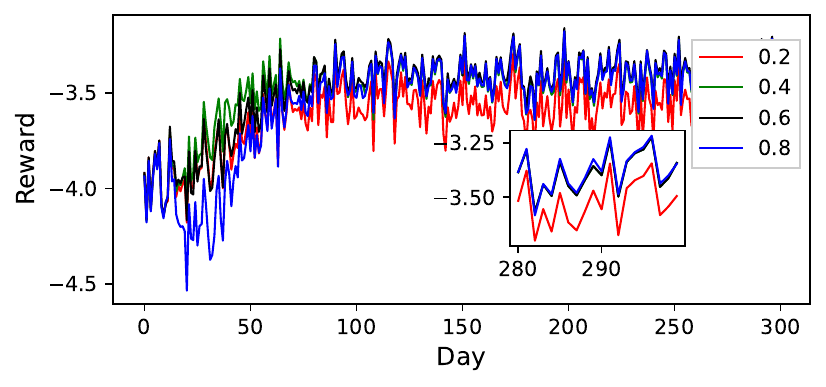}
\end{minipage}
    }
\subfigure[118-bus]{
\begin{minipage}[t]{0.315\linewidth}\label{wm1_large_sacle_118}
\includegraphics[width=2.4in]{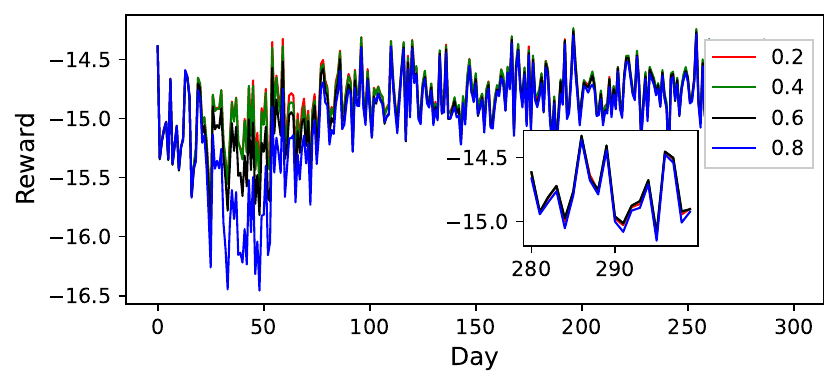}
\end{minipage}
    }
\caption{Testing results of the training stage for 33-bus, 69-bus and 118-bus distribution networks.} \label{wm1_large_sacle}
\end{figure*}


\begin{figure*}[ht]\label{wm2_large_sacle}
\centering
\subfigure[33-bus]{\begin{minipage}[t]{0.315\linewidth}\label{wm2_large_sacle_33}
    \includegraphics[width=2.4in]{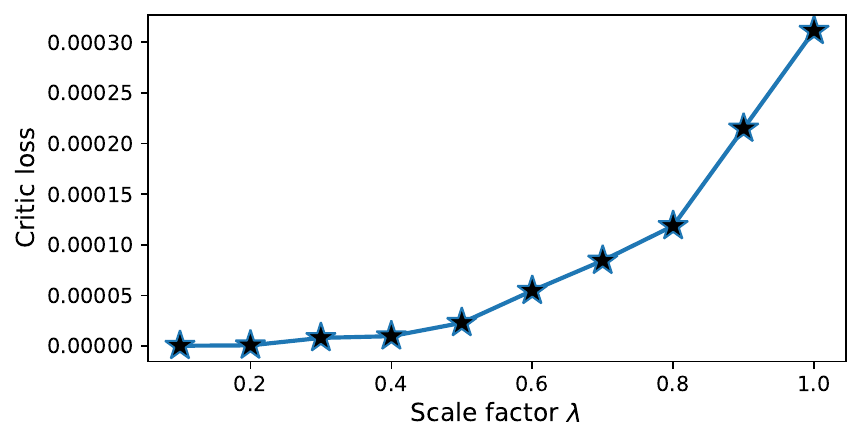}
    \end{minipage}
    }
\subfigure[69-bus]{
\begin{minipage}[t]{0.315\linewidth}\label{wm2_large_sacle_69}
\includegraphics[width=2.4in]{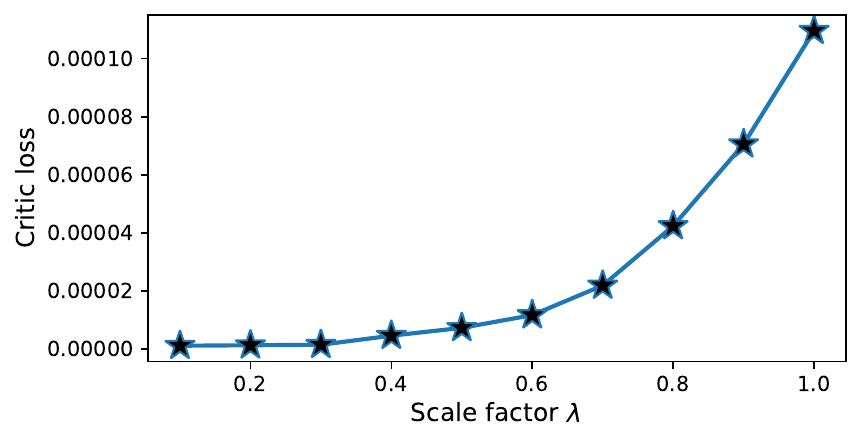}
\end{minipage}
    }
\subfigure[118-bus]{
\begin{minipage}[t]{0.315\linewidth}\label{wm2_large_sacle_118}
\includegraphics[width=2.4in]{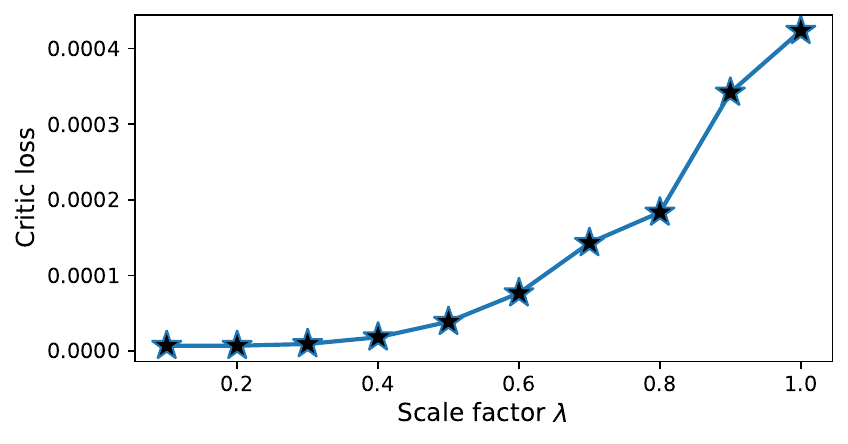}
\end{minipage}
    } 
\caption{The change of the critic loss with increasing residual action space in the final 50 days.}\label{wm2_large_sacle}
\end{figure*}

\begin{figure*}[ht]
\centering
\subfigure[33-bus]{\begin{minipage}[t]{0.315\linewidth}\label{wm1_large_sacle_33}
    \includegraphics[width=2.4in]{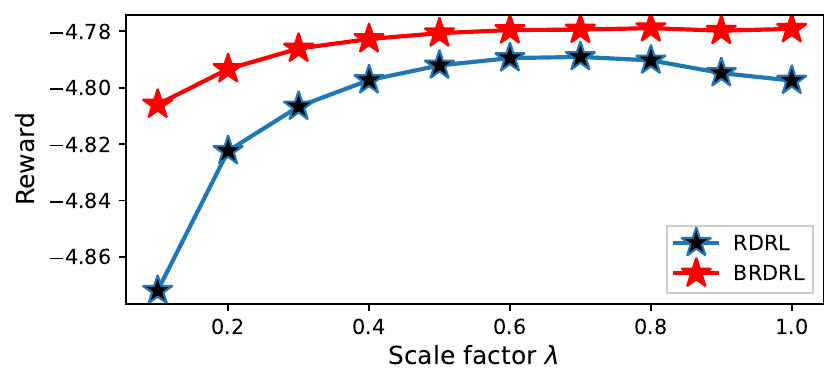}
    \end{minipage}
    }
\subfigure[69-bus]{
\begin{minipage}[t]{0.315\linewidth}\label{wm3_large_sacle_69}
\includegraphics[width=2.4in]{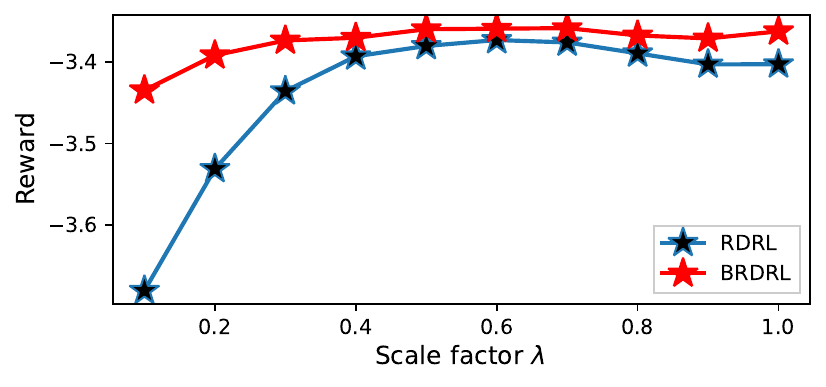}
\end{minipage}
    }
\subfigure[118-bus]{
\begin{minipage}[t]{0.315\linewidth}\label{wm3_large_sacle_118}
\includegraphics[width=2.4in]{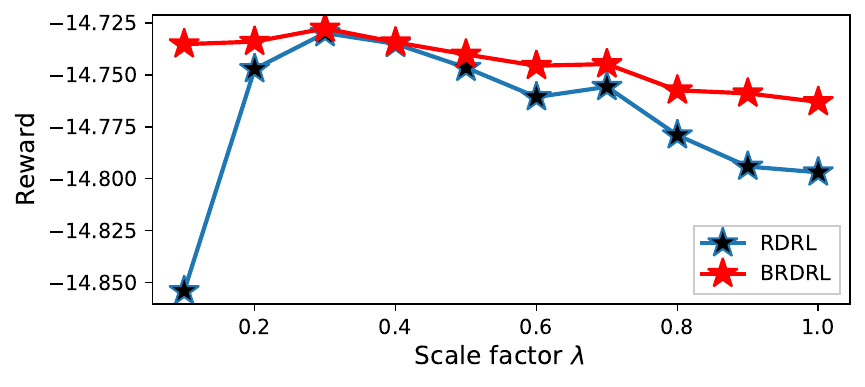}
\end{minipage}
    }
\caption{The change of the reward with the increasing residual action space in the final 50 days.
RDRL is residual DRL, and the results were from simulation 4). BRDRL is boosting residual DRL, and the results were from simulation 5).
} \label{iteration}
\end{figure*}

\textbf{Residual policy learning:}
We abbreviate RDRL with $\lambda =1$ as "RDRL-E" because the size of its residual action space is equal to the corresponding SAC.
Comparing the results of RDRL-E with DRL showed the advantages of "residual policy learning". 
As shown in Fig. \ref{RDRL_E_trajectory}, in the initial learning stage (days 10-50), even though there was a slight fluctuation in the initial training stage, RDRL-E achieved considerably better than DRL.
As shown in Fig. \ref{RDRL_E}, after enough time to learn, residual policy learning reduced the reward error $28\%,  15\%, 55\%$ in 33, 69, and 118 bus distribution networks.  The value is calculated by the reward of (DRL - RDRL-E)$/$DRL.
Those verified residual policy learning reduced the learning difficulties of the actor and improved the optimization performance of RDRL.

\textbf{Learning in a reduced residual action space:} Comparing the results of RDRL with RDRL-E shows the advantages of ``learning in a reduced action space".
As shown in Fig. \ref{RDRL_E_trajectory}, in the initial learning stage (days 10-50), the slight fluctuation is invisible for RDRL, and RDRL achieved considerably better than DRL-E.
As shown in Fig. \ref{RDRL_E}, after enough time to learn, ``learning in a reduced action space" further declined the reward error $16\%,  34\%, 20\%$ in 33, 69, and 118 bus distribution networks. The value is calculated by the reward of (RDRL-E - RDRL)$/$DRL.
Those verified learning in a reduced residual action space reduced the learning difficulties of the actor and improved the optimization performance of RDRL.

As we discuss, there are two rationales for the effectiveness of learning in a reduced residual action space. They were further demonstrated by the following simulation phenomenons:
\begin{itemize}
    \item[1)] As shown in Fig. \ref{wm1_large_sacle}, during the initial learning stage (days 10-50), smaller residual action spaces lead to smaller fluctuations in learning trajectories. For clear presentation, Fig. \ref{wm1_large_sacle} only shows the learning trajectories of reward with the scale factor for residual action space $\lambda = 0.2, 0.4, 0.6, 0.8$.
    \item[2)] As shown in Fig. \ref{wm2_large_sacle}, small residual action space leads to small critic loss.  
    Fig. \ref{wm2_large_sacle} shows the 10 experiment results where the critic loss changes with increasing residual action space during the final 50 days.
    The critic loss was the means of critic error for the sampling batch data in the final 50 days. 
    The critic with smaller errors provides a more accurate guidance for actor network.
\end{itemize}

\subsection{ BRDRL alleviates the ``too small" or ``too large" problems of RDRL}

As we discussed, the size of the residual action space is one of the crucial reasons for improving the RDRL performance. The residual action space is a tunable parameter in RDRL. The unsuitable setting residual action space would degrade the optimization performance.
It was demonstrated by simulation 4) that the scale factor of residual action space is $\lambda = 0.1,0.2,0.3,\dots 1$. The change of the testing reward with the increasing residual action space in the final 50 days is shown in  Fig. \ref{iteration}.
It shows the issues of ``too small" or ``too large" residual action space of RDRL and the benefits of BRDRL. Here are the detailed discussions:
\begin{itemize}
    \item[1)] \textbf{``Too small" residual action space cannot cover the optimal action:}
    With the increase of residual action space, the reward of RDRL first increased but then decreased.
    For $\lambda = 0.1,\dots, 0.6$ in 33 and 69 bus distribution network, and $\lambda = 0.1,0.2, 0.3$ in 118-bus distribution network, rewards increasing with the increasing of action space indicated the problem of ``too small"  residual action space. At this stage, the residual action space cannot cover the optimal action and the final actions cannot reach the optimal actions. Increasing the action space alleviated the problem and made the actions more likely to close optimal actions.
    \item[2)] \textbf{``Too large" residual action space degrades the optimization performance:}
    As shown in the reward trajectory labeled ``RDRL" of $\lambda =0.6,\dots, 1$ for 33 and 69 bus distribution network, and $\lambda = 0.3,\dots,1$ for 118 bus distribution network, the residual action space increased and the rewards decreased.
    It indicated the problem of ``too large"  action space.
    The ``too large" residual action space increased the learning difficulties of DRL, thus leading to a declined optimization performance.
\end{itemize}

BRDRL is an effective approach to alleviate the ``too small"  or ``too large"  problem of residual action space.
It was verified by simulation 5), and the results were shown Fig. \ref{iteration}.
For the ``too small" residual action space of RDRL, BRDRL learned the next residual action with $\lambda = 0.2$. 
The residual action approached the optimal value further, thus improving the optimization performance. 
For the ``too large" residual action space, BRDRL learned on a much smaller residual action space with $\lambda = 0.2$, further improving the optimization capabilities.

\section{Conclusion}
This paper proposed RDRL that learns the residual action of the base action from the model-based optimization under an approximate power flow model. It improved the DRL performance throughout the whole training process by inheriting the control capability of the model-based optimization, residual policy learning, and learning in smaller residual action space. 
Meanwhile, we found that the ``too small" or ``too large" residual action space degraded the RDRL performance. To alleviate the two problems and improve the performance further, we extend the RDRL to the BRDRL. 
Corresponding simulations verified the superiority of the proposed RDRL and BRDRL. We also designed simulations to verify the three rationales of the effectiveness of RDRL, and BRDRL alleviated the “too small” or “too large” problems of RDRL point-by-point. 

The proposed method is a general and comprehensive method for constrained optimization problems.
In the future, we will extend the method to more optimization problems in the power system field to achieve more desirable outcomes in real-world engineering conditions.

\ifCLASSOPTIONcaptionsoff
  \newpage
\fi



\bibliographystyle{IEEEtran}
\bibliography{IEEEabrv,My_Library.bib}
\end{document}